\documentclass[pra,twocolumn,amsmath,showpacs]{revtex4}
\usepackage{graphics}

\newcommand{\hpsi}{\ensuremath{\hat{\Psi}(x)}}
\newcommand{\hpsid}{\ensuremath{\hat{\Psi}^\dagger(x)}}
\newcommand{\dhpsi}{\ensuremath{\delta\hpsi}}
\newcommand{\dhpsid}{\ensuremath{\delta\hpsid}}
\newcommand{\phizero}{\ensuremath{\phi_0(x)}}
\newcommand{\hbk}{\ensuremath{\hat{b}_k}}
\newcommand{\hbkd}{\ensuremath{\hat{b}^\dagger_k}}
\newcommand{\hok}{\ensuremath{\hbar\omega_k}}
\newcommand{\uk}{\ensuremath{u_k(x)}}
\newcommand{\uks}{\ensuremath{u_k^{*}(x)}}
\newcommand{\vk}{\ensuremath{v_k(x)}}
\newcommand{\vks}{\ensuremath{v_k^{*}(x)}}
\newcommand{\uek}{\ensuremath{u^{(e)}_k(x)}}
\newcommand{\uok}{\ensuremath{u^{(o)}_k(x)}}
\newcommand{\gno}{\ensuremath{gn_0(x)}}
\newcommand{\dek}{\ensuremath{\delta_e(k)}}
\newcommand{\dok}{\ensuremath{\delta_o(k)}}

\newcommand{\lambdat}{\ensuremath{\tilde{\lambda}}}

\begin{document}
\title{Scattering of atoms on a {B}ose--{E}instein condensate}
\author{Uffe V. Poulsen} 
\email{uvp@ifa.au.dk} 
\author{Klaus M{\o}lmer} 
\affiliation{ QUANTOP, Department of Physics and
  Astronomy, University of Aarhus, DK-8000 \AA rhus C}
\begin{abstract}
  We study the scattering properties of a Bose--Einstein condensate
  held in a finite depth well when the incoming particles are
  identical to the ones in the condensate. We calculate phase shifts
  and corresponding transmission and reflection coefficients, and we
  show that the transmission times can be negative, i.e., the atomic
  wavepacket seemingly leaves the condensate before it arrives.
\end{abstract}
\pacs{3.75.Fi,34.50-s}
\maketitle
\section{Introduction}
\label{sec:introduction}

In this paper we present an analysis of the scattering of atoms on a
Bose-Einstein condensate. A repulsive interaction between atoms is
both responsible for the shape of the condensate and for the
interaction between the incident atom and the condensed atoms.
Scattering effects can be significant when two condensates merge, and
four-wave mixing has been observed in experiments
\cite{deng99:_four}.  We deal here with the limit of a
single or just a few atoms, incident in a well defined momentum state
on a condensate, a situation that has been realized experimentally to
seed atom lasers \cite{vogels02:_gener_bose}
and 'atomic parametric amplifiers'
\cite{inouye99:_phase,inouye00:_amplif_bose} with a weak
atomic beam. Our interest, however, is in the interaction between the
small atomic component and the condensate, and in particular in the
transmission and reflection properties of the condensate as a beam
splitter for atoms of the same kind as the mirror itself. Since the
atoms of the scatterer are indistinguishable from the scattered
particles, exchange effects play an important role, and similar
scattering studies have indeed been proposed as a means for
investigating superfluidity in strongly interacting He-4 system
\cite{halley93:_new,setty97:_variat_monte_carlo_studies_rate}, and
also in weakly interacting systems
\cite{wynveen00:_ident_bose_einst}.

Dilute condensates can be trapped in a large number of trapping
arrangements, and we shall assume a trapping potential
of finite width allowing asymptotically free atoms to be directed
towards the trapped condensate and interact with it in a well defined
region of space. Such potentials can be created optically or possibly
in the region above a current carrying chip, and our calculations,
which will be carried out in a one dimensional geometry, may
correspond to either a condensate slab hit by atoms with well defined
transverse momenta or to a trapped condensate in a wave guide with a 
local longitudinal minimum.

In Sec.~\ref{sec:bogoliubov-approach} we introduce the Bogoliubov
treatment of excitations of a Bose-Einstein condensate. In
Sec.~\ref{sec:results} we explain how an analysis of the fundamental
excitations of the condensate provides the scattering information.
Numerical results are compared with an approximate analytical model.
In Sec.~\ref{sec:td-scat}, we show examples of time dependent
wavepacket dynamics, and we identify a momentum regime in which
transmission occurs with negative time delays, i.e., the wave packet
emerges from the condensate before it arrives.
Sec.~\ref{sec:conclusion} concludes the paper.

\section{The Bogoliubov Approach}
\label{sec:bogoliubov-approach}
In the Bogoliubov approach, the number of particles in the condensate
mode is assumed to be high and rather well defined, and the problem of
small thermal or externally induced excitations can be treated to
lowest order to yield a picture of non-interacting quasi-particles.
These are the normal modes of the system and they are mixtures of
particle- and hole-like excitations.  In the case of scattering, the
incoming particle will behave as a quasiparticle while moving through
the condensate, whereas asymptotically the excitation is particle-like
as hole-excitations are limited to the region occupied by the
condensate.

Let us briefly review the Bogoliubov approach. First, the field
operator is written in a form explicitly emphasizing the condensate
mode
\begin{equation}
  \label{eq:splitfield}
  \hpsi = \sqrt{N} \phizero + \dhpsi.
\end{equation}
Here $\phizero$ is to be regarded as a $c$-number field describing the
condensate while $\dhpsi$ describes the correction. We insert this
form in the full Hamiltonian (in the contact--interaction
approximation),
\begin{multline}
  \label{eq:fullhamilton}
  \hat{H}=
  \int\!dx\; \biggl\{
    \hpsid h(x) \hpsi
    \\
    +
    \frac{g_\text{1D}}{2} \hpsid \hpsid \hpsi \hpsi
  \biggr\},
\end{multline}
where $h(x)=-(\hbar^2/2m)\partial_x^2+V(x)$ is the single-particle
Hamiltonian and $g_\text{1D}$ quantifies the 1D interaction strength.
Keeping only interaction terms of order $N$ or higher, we arrive at a
quadratic form in $\dhpsi$ and $\dhpsid$. The terms linear in $\hpsi$
and $\hpsid$ are identically zero if $\phizero$ solves the
Gross--Pitaevskii equation
\begin{equation}
  \label{eq:gpe}
  \left[
    h(x) 
    +
    g N |\phizero|^2
  \right]
  \phizero
  =
  \mu \phizero.
\end{equation}
To diagonalize the remaining part, one envokes a Bogoliubov
transformation and writes
\begin{equation}
  \label{eq:bogo_trans}
  \dhpsi = 
  \sum_{k} \left[\hbk \uk - \hbkd \vks\right]
  ,
\end{equation}
where $\uk$ and $\vk$ solve the Bogoliubov--de Gennes equations:
\begin{align}
  \label{eq:bog-degen_eq1}
  [ L(x) - \hok ] \uk & = \gno \vk \\
  \label{eq:bog-degen_eq2}
  [ L(x) + \hok ] \vk & = \gno \uk 
\end{align}
and are normalized to 
\begin{equation}
  \label{eq:bog_norm}
  \int \left[ \uks u_{k'}(x) - \vks v_{k'}(x) \right]dx
  =
  \delta_{kk'}.
\end{equation}
The operator $L(x)$ is defined by $L(x)=h(x)+2\gno-\mu$.  The
Hamiltonian finally takes the simple form
\begin{equation}
  \label{eq:H-diag}
  \hat{H}_\text{diag} = E(N)+\sum_{k=1}^{\infty} \hok \hbkd \hbk,
\end{equation}
i.e, it describes non-interacting quasi--particles created by $\hbkd$
and destroyed by $\hbk$. Note how the Bogoliubov transform mixes
creation and destruction operators. This means, that excitations have
both a particle--like character (described by $u_k$) and a hole--like
character (describe by $v_k$).

The Bogoliubov analysis is routinely applied to the case of a
homogeneous situation (flat potential) and the case of an infinite
trapping potential~\cite{dalfovo99:_theor_bose_einst}. In the first
case, one obtains solutions with well--defined momenta: phonons in the
long wave--length limit, free particles for high momenta. The spectrum
is continuous and gap--less.  In the second case, the excitations are
again collective in the low energy regime, approaching
single--particle trap states for high energies~\cite{Ohberg1997a}. The
spectrum is discrete. Here we will concentrate on a third situation
where the trapping potential has finite width and depth. In this
scenario, a finite number of trapped excitations exists and above
these a continuum of scattering states. The corresponding spherically
symmetric problem in 3D has been treated by Wynveen et al.\ in
\cite{wynveen00:_ident_bose_einst}.

Away from the trap and thus from the condensate, the scattering states
are particle--like ($\vk=0$) and the usual asymptotic analysis of
scattering applies: We can define incoming, reflected, and transmitted
fluxes and we can set up an initial wave--packet and make it progate
towards the condensate region. The actual tranmission and reflection
coefficients depend on what happens while the incoming particle passes
through the condensate and this propagation is partly phonon--like,
i.e., with the emergence of a hole--like component.

\begin{figure}[tbp]
  \includegraphics{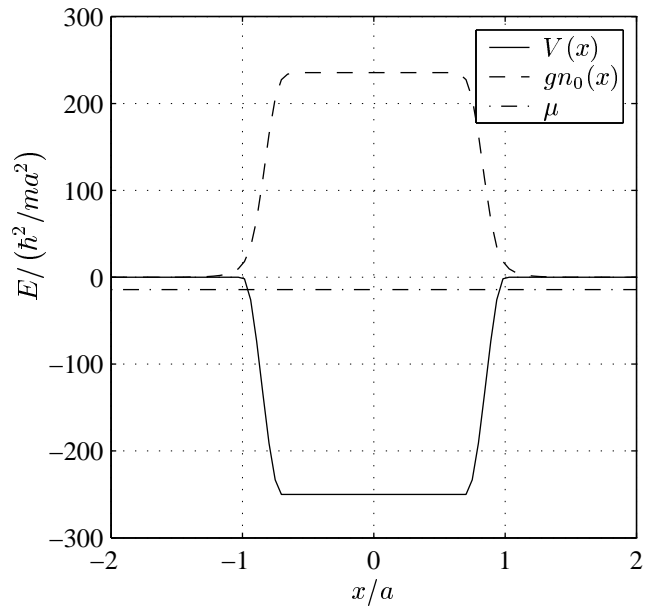}
  \caption{Figure showing the potential well and the shape of the
    condenste. The range of the potential is $[-a,a]$ outside which
    $V(x)$ is identically 0. $V(x)$ changes smoothly to
    $-V_0=-250\hbar^2/ma^2$ in edge--zones of width $0.3a$. The 1D
    interaction strength is taken to be $g_\text{1D}N=400\hbar^2/ma$. The
    solution of the GPE~(\ref{eq:gpe}) then gives
    $\mu=-14.4\hbar^2/ma^2$ and a Bogoliubov speed of sound
    $c=\sqrt{gn_0(0)/m}=15.4\hbar/ma$ in the inner region of the well.}
  \label{fig:well}
\end{figure}

In 1D, the Bogoliubov--De Gennes equations are computationally quite
manageable and the choice of method for their solution is not crucial.
Nevertheless, for completeness we here briefly summarize our method.
First we solve the Gross--Pitaevskii equation (\ref{eq:gpe}) with the
chosen potential. We do this by a simple steepest descent method,
i.e., by propagating the corresponding time--dependent equation in
imaginary time. In particular, we use a split--step
fast--Fourier--transform algoritm~\cite{fleck76:_time-dep_prop}.
In Fig.~\ref{fig:well}, we show the particular potential we will be
using, and we show the corresponding condensate wavefunction. 

When $\phizero$ and $\mu$ have been found,
Eqs.~(\ref{eq:bog-degen_eq1}) and (\ref{eq:bog-degen_eq2}) are
solved by first defining
\begin{equation}
  \label{eq:def_fh}
  \begin{split}
      f_k(x)&=\sqrt{\frac{1}{2}}\left[\uk+\vk\right]
      \\
      h_k(x)&=\sqrt{\frac{1}{2}}\left[\uk-\vk\right]
  \end{split}
\end{equation}
in terms of which
Eqs.~(\ref{eq:bog-degen_eq1}) and (\ref{eq:bog-degen_eq2}) become
\begin{align}
  \label{eq:bg_hf1}
    [ L(x)-\gno] f_k(x) & = {\hok} h_k(x)\\
    \label{eq:bg_hf2}
    [ L(x)+\gno] h_k(x) & = \hok f_k(x) 
\end{align}
We can then avoid diagonalizing a two--component problem by simply
applying the operator $[L(x)+\gno]$ to both sides of
Eq.~(\ref{eq:bg_hf1}) to find the necessary condition
\begin{equation}
  \label{eq:fourth_order}
    [ L(x)+\gno][ L(x)-\gno] f_k(x)  =  [\hok]^2 f_k(x). 
\end{equation}
When this one--component eigenvalue problem is solved, $\uk$ and $\vk$
can be found by applying first Eq.~(\ref{eq:bg_hf1}) and then
Eq.~(\ref{eq:def_fh}).
 
\section{Results}
\label{sec:results}

\subsection{Phaseshifts}
\label{sec:phaseshifts}
We assume that $V(x)$ and thus $n_0(x)$ have even spatial symmetry. We
can then demand that solutions of Eqs.~(\ref{eq:bog-degen_eq1}) and
(\ref{eq:bog-degen_eq2}) have definite parity. In the discrete part of the
spectrum, even and odd solutions alternate, while in the continuum,
each energy supports solutions of both parities. In the asymptotic region
away from the potential and the condensate, even and odd scattering
solutions can then be written
\begin{equation}
  \label{eq:asymp_solu}
  \begin{array}[c]{lcl}
    \uek &\rightarrow &
    \cos(kx\mp\dek) \\
    \uok &\rightarrow &
    \sin(kx\mp\dok)
  \end{array}
  \quad \text{for\ } x\rightarrow \pm \infty
  . 
\end{equation}
Note that we identify the label $k$ with the asymptotic wavenumber for
the scattering solutions. The phaseshifts, $\dek$ and $\dok$, contain
all the information about the scattering relevant to the asymptotic
region. They can be conveniently extracted from our numerical
solutions for $u_k$. The calculations are actually done on a
space--interval $[-L,L]$ with periodic boundary conditions (finite
lower momentum cutoff). This means that for a given $L$ we only find the
subset of the continuum solutions with wavenumbers
fulfilling
\begin{equation}
  \label{eq:boundary}
  -kL+\delta(k) \equiv kL-\delta(k) \mod 2\pi
\end{equation}
or, equivalently,
\begin{equation}
  \label{eq:delta_kL}
  \delta(k) \equiv kL \mod \pi.
\end{equation}
After the diagonalization this discrete set of $k$ values can easily
be determined from the eigenvalues as $\hok=\hbar^2 k^2/2m - \mu$.
Each $k$ value gives us one point on $\delta(k)$ via
Eq.~(\ref{eq:delta_kL}). This is enough to determine $\delta(k)$ if it
is slowly varying on the $1/L$ scale. If not, we just need to change
$L$ slightly and repeat the calculation to obtain an additional set of
points.

\begin{figure}[tbp]
  \includegraphics{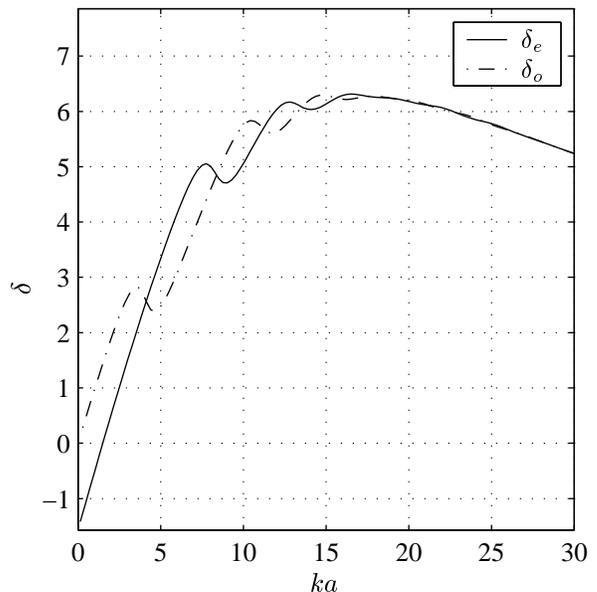}
  \caption{Phaseshifts of even and odd solutions as a function of
    incoming momentum. Parameters as in Fig.~\ref{fig:well}.}
  \label{fig:deltas}
\end{figure}

In Fig.~\ref{fig:deltas} 
we show a typical example of $\dek$ and
$\dok$ with parameters as in Fig.~\ref{fig:well}. There is clearly
some resonant behavior with out--of--phase oscillations of $\dek$ and
$\dok$ around a slowly varying average. This behaviour is well known
from ordinary, single particle scattering on a well/barrier, and in the
following subsection we shall present an analytical model which yields
the same gross features.

\subsection{Square well model, Thomas--Fermi approximation}
\label{sec:square-well}

Let us consider a condensate trapped in a square well potential 
\begin{equation}
  \label{eq:square_well}
  V(x)
  =
  \left\{
    \begin{array}[c]{lcl}
      -V_0 & , & |x|< a_\text{sw} \\
      0 & , & |x|> a_\text{sw} \\
    \end{array}
  \right.
\end{equation}
where $a_\text{sw}$ is an appropriately defined effective width.
In the Thomas--Fermi approximation the condensate wave function is
constant in the trap and zero outside
\begin{equation}
  \label{eq:square_cond}
  \phizero
  =
  \left\{
    \begin{array}[c]{lcl}
      \sqrt{\frac{N}{2a_\text{sw}}} & , & |x|< a_\text{sw} \\
      0 & , & |x|> a_\text{sw} \\
    \end{array}
  \right.
  .
\end{equation}
This $\phizero$ is naturally not a solution to the Gross--Pitaevskii
equation as the kinetic energy term will smoothen the step in density
even when the external potential is discontinuous. Formally, this
introduces linear terms in Eq.~(\ref{eq:H-diag}), i.e., the Bogoliubov
vacuum is not a steady--state for the system. We are, however, only
interested in the scattering behaviour at positive energies and it is
reasonable to assume that some insight can be gained by finding
solutions to Eqs.~(\ref{eq:bog-degen_eq1}) and
(\ref{eq:bog-degen_eq2}) with the simplifying assumptions expressed by
(\ref{eq:square_well}) and (\ref{eq:square_cond}).

It is amusing to note, that we are now dealing with the Bogoliubov--de
Gennes analog of the undergraduate textbook problem of 1D scattering
on a square well. The full solution is found by matching the
analytical wavefunctions in the regions, $x\in
[-\infty,-a_\text{SW}]$, $[-a_\text{SW},a_\text{SW}]$, and
$[a_\text{SW},\infty]$. To the left and to the right of the well, we
demand $v(x)=0$ and let $u(x)=\cos(kx\pm\dek)$
($u(x)=\sin(kx\pm\dok)$) to find even (odd) solutions. Inside the
well, we are in a region of constant potential ($V(x)=-V_0$) and
constant condensate density ($N|\phizero|^2=n_0$) so
Eqs.~(\ref{eq:bog-degen_eq1}) and (\ref{eq:bog-degen_eq2}) read
\begin{align}
  \label{eq:BG_flat1}
  \left[ 
    -\frac{\hbar^2}{2m}\partial_x^2-V_0+2gn_0-\mu - \hbar\omega 
  \right] u(x) 
  & = gn_0 v(x) \\
  \label{eq:BG_flat2}
  \left[ 
    -\frac{\hbar^2}{2m}\partial_x^2-V_0+2gn_0-\mu + \hbar\omega 
  \right] v(x)
  & = gn_0 u(x)
  .
\end{align}
To be able to match the boundary conditions, we need four
linearly independent solutions and the ansatz $u(x)=U e^{i\lambda x}$,
$v(x)=V e^{i\lambda x}$ leads to $\lambda \in \{\pm \lambda, \pm
\lambdat \}$ where
\begin{multline}
  \label{eq:lambda}
  \hbar\lambda(k) 
  = 
  \\
  \sqrt{
    2m\left(
      \sqrt{
        \left[
          gn_0
        \right]^2
        +
        \left[
          E-\mu
        \right]^2
        }
      -
      2gn_0+V_0+\mu
    \right)
    }
\end{multline}
and
\begin{multline}
  \label{eq:lambdat}
  \hbar\lambdat(k)
  =
  \\
  i\sqrt{
    2m\left(
      \sqrt{
        \left[
          gn_0 
        \right]^2
        +
        \left[
          E-\mu
        \right]^2
        }
      +
      2gn_0-V_0-\mu
    \right)
    }
\end{multline}
where $E=(\hbar k)^2/2m$ is the incoming kinetic energy. Note that
$\lambdat$ is imaginary and it is normally disregarded in homogeneous
condensates. Here, it is needed as we must match the boundary
conditions that also $v$ and $v'$ are continuous.

The matching of $u$ and $v$ at $x=a_\text{sw}$ leads to equations 
for the phaseshifts. They read
\begin{eqnarray}
  \label{eq:dek_match}
  \tan\left(
    ka_\text{sw}-\dek
  \right)
  &=&
  \frac{\lambda(k)}{k}
  \tan\left(
    \lambda(k)a_\text{sw}
  \right)
  \\
  \label{eq:dok_match}
  \tan\left(
    ka_\text{sw}-\dok
  \right)
  &=&
  \frac{k}{\lambda(k)}
  \tan\left(
    \lambda(k)a_\text{sw}
  \right)
  .
\end{eqnarray}
These equations are the same as for single particle square well
scattering except that the usual
$\hbar\lambda=\sqrt{2m(E-V)}$ is replaced by $\hbar\lambda$
from Eq.~(\ref{eq:lambda}).   

\begin{figure}[tbp]
  \includegraphics{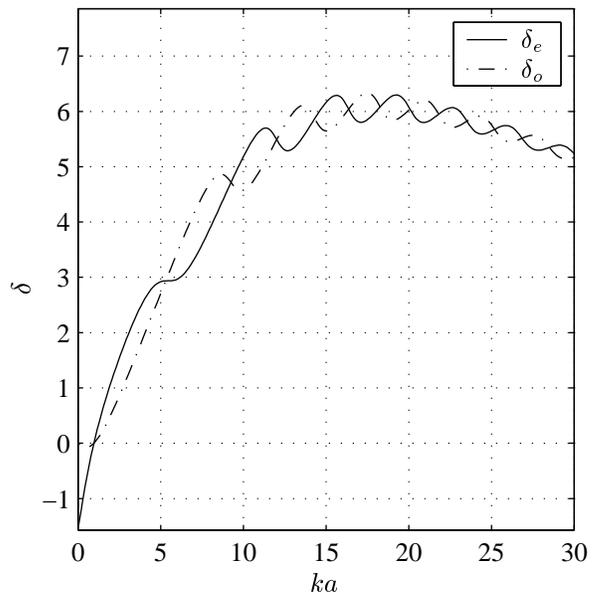}
  \caption{Even and odd phaseshifts for a simple square well/square  condensate
    model. The parameters are like in Fig.~\ref{fig:well} and the
    curves should be compared to the full numerical calculations
    presented in Fig.~\ref{fig:deltas}. There is clearly a qualitative
    agreement: The smooth behavior of the average of the
    two curves is well reproduced and the periode of the oscillations
    around this average is also of the correct order of
    magnitude. However, the oscillations extend to too high values of
    $k$.} 
  \label{fig:square}
\end{figure}

In Fig.~\ref{fig:square} we show results obtained for parameters like
in Fig.~\ref{fig:well}: $gn_0$ is taken as the central density in the
smooth trap and $a_\text{sw}$ is chosen to accomodate the same total
number of particles. A comparison with Fig.~\ref{fig:deltas} reveals
both similarities and differences: The smooth behaviour of the average
of the two curves is well reproduced as well as the (quasi--)period of
the oscillatory behaviour. However, the positions of curve crossings
and the maxima of the phaseshift differences, especially at low $k$'s,
are not well reproduced. Also, at high momenta the sharp--edge
approximation leads to stronger oscillations in $\dek$ and $\dok$ than
for the smooth well. 

\subsection{Transmission coefficent}
\label{sec:transm-coeff}
\begin{figure}[tbp]
  \includegraphics{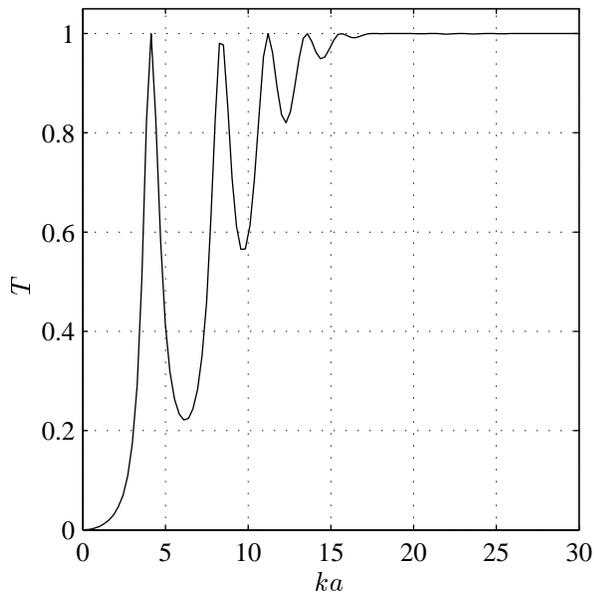}
  \caption{\label{fig:RT}Transmission and reflection coefficients for
    the situation considered in Figs.~\ref{fig:well} and
    \ref{fig:deltas}. A series of perfect transmissions are seen,
    corresponding to the crossings of the $\dek$ and $\dok$ curves in
    Fig.~\ref{fig:deltas}. }
\end{figure}

The typical scattering situation with an incoming, a reflected, and a
transmitted wave specifies the asymptotic form of the wave function 
\begin{equation}
  \label{eq:scat_uk}
  u_k(x) \rightarrow 
  \left\{
    \begin{array}[c]{lcl}
      e^{ikx}+R(k)e^{-ikx} & \text{for} & x \rightarrow -\infty \\ 
      T(k)e^{ikx} & \text{for} & x \rightarrow \infty 
    \end{array}
  \right.
  .
\end{equation}
This defines the reflection- and transmission-coefficients $R(k)$ and
$T(k)$. Making the change of basis from the odd and even parity
eigenstates~(\ref{eq:asymp_solu}), one finds
\begin{align}
  \begin{split}
    \label{eq:R_of_delta}
    R(k)
    &=
    \frac{1}{2}\left(e^{-2i\dek}-e^{-2i\dok}\right)\\ 
    &=
    ie^{-i(\dok+\dek)}\sin(\dok-\dek)
  \end{split}
  \\
  \begin{split}
    \label{eq:T_of_delta}
    T(k)
    &=
    \frac{1}{2}\left(e^{-2i\dek}+e^{-2i\dok}\right)\\
    &=
    e^{-i(\dok+\dek)}\cos(\dok-\dek)
    .
  \end{split}
\end{align}
We see that 100\% transmission takes place at $k$'s where $\dek=\dok$
while 100\% reflection requires $\dek=\dok\pm\pi/2$. In
Fig.~\ref{fig:RT} we plot $R$ and $T$ for the same situation as
considered above. The curve crossings in Fig.~\ref{fig:deltas} are now
translated into transmission windows. At very low
momenta, we get 100\% reflection and at very high momenta we
naturally get 100\% transmission.  

\section{Time dependent scattering}
\label{sec:td-scat}

\subsection{Wavepacket dynamics}
\label{sec:wavepacket-dynamics}

With the complete set of scattering states it is possible to follow
scattering of wavepackets in time. To obtain the initial state we add
a number of particles to the Bogoliubov vacuum:
\begin{equation}
  \label{eq:initial_state}
  |t=0\rangle
  =
  \frac{1}{\sqrt{n_s!}}
  \left[ 
    \int \phi_\text{s}(x) \dhpsid 
  \right]^{n_s}
  | \text{vac} \rangle
\end{equation}
where $\phi_\text{s}(x)$ is the desired wavepacket mode. The operator
term in this equation creates an $n_s$--particle state, but to benefit
from the simple form of Eq.~(\ref{eq:H-diag}) we should think of it
as creating $n_s$ quasiparticles which at $t=0$ just happen to be
localized well away from the condensate. The quasiparticles are
created in a superposition of energy eigenstates and the coefficients
in this superposition are found as
\begin{equation}
  \label{eq:def_ck}
  c_k
  =
  \int \left[
    \uks \phi_\text{s}(x) + \vks \phi_\text{s}^*(x)
  \right] dx
  .
\end{equation}
As $\phi_s$ is located well away from the condensate region, there is
in fact no contribution from the $v$ part of this integral.

The time evolution is entirely given by the relation
$\hbk(t)=e^{-i\omega_k t}\hbk$ in the Heisenberg picture, and we find
the non--condensate mode part of the total density to be
\begin{equation}
  \label{eq:out_of_condensate}
  \langle \dhpsid \dhpsi \rangle
  =
  n_s 
  \left|
    U(x,t)
  \right|^2
  +
  n_s
  \left|
    V(x,t)
  \right|^2
  +
  \sum_k \left| \vk \right|^2
\end{equation}
where
\begin{align}
  \label{eq:def_UV}
  U(x,t)&=\sum_k c_k e^{-i\omega_k t} \uk\\
  V(x,t)&=\sum_k c_k e^{-i\omega_k t} \vk
  .
\end{align}
The last term in Eq.~(\ref{eq:out_of_condensate}) is the quantum
depletion, which is always present due to the interactions in the
condensate, while the first two terms are consequences of the
scattering process.

Depending on the spread of $k$-values in the wavepacket, the
asymptotic behaviour of the scattering will be more or less simply
described by reading off the reflection and transmission coefficients
$R$ and $T$ at the average momentum $k_0$.  In fact, looking at the
first form of these coefficients in
Eqs.~(\ref{eq:R_of_delta}) and (\ref{eq:T_of_delta}) we are reminded that
the reflected and the transmitted wave can be seen as superpositions
of an even part and an odd part. To each of these can be ascribed a
time-delay, $\Delta t_{e/o}$ in the arrival of the original wave
packet at certain point in space with respect to the free propagation.
It is easy to show that these delays are given by
\begin{eqnarray}
  \label{eq:delay_even}
  \Delta t_{e}
  &=&
  \left.
    -\frac{2}{v_g}\frac{\partial \dek}{\partial k}
  \right|_{k=k_0}
  \\
  \label{eq:delay_odd}
  \Delta t_{o}
  &=&
  \left.
    -\frac{2}{v_g}\frac{\partial \dok}{\partial k}
  \right|_{k=k_0}
\end{eqnarray}
where $v_g = \hbar k_0/m$ is the (group) velocity of the incoming
wavepacket. If these delays are sufficiently different the reflected
and transmitted wavepackets are expected to be doublepeaked. In fact,
a closer look at Eqs.~(\ref{eq:R_of_delta}) and~(\ref{eq:delay_odd})
reveals a difference in sign: the reflected wavepacket can easily be
double--peaked as it is the difference between the even and odd
contribution, while the transmitted wavepacket is the sum and
therefore require displacements as large as the wavepacket width
for a visible effect.

\begin{figure}[tbp]
  \includegraphics{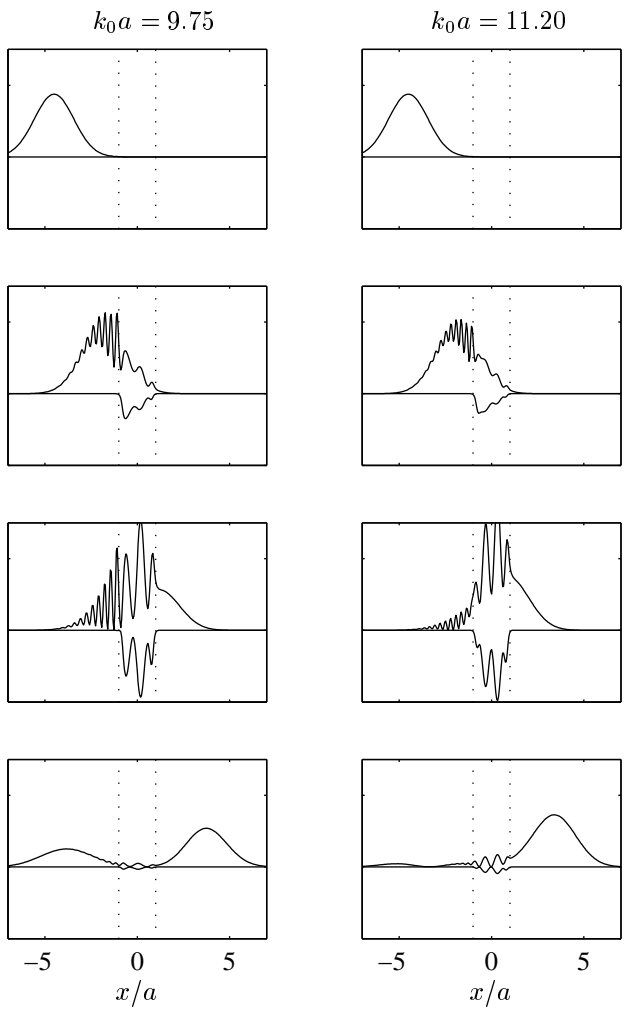}
  \caption{Wavepacket scattering on a BEC. Time series are shown for
    two different incoming momenta: At mean wavenumber $k_0a=9.75$, the
    initial packet propagates towards the potential well containing 
    the BEC and reflected and transmitted wavepackets appear. In
    accordance with Fig.~\ref{fig:RT} transmission is approximately
    62\% at this $k_0$. At $k_0a=11.2$ we are in a transmission window
    and the reflected wavepacket has small amplitude and a
    double--peaked envelope.The figures show both $|U(x)|^2$ and
    $-|V(x)|^2$, the particle contribution and ($-$) the hole contribution
    to the total density of scattering atoms. Note that the ground
    state quantum depletion also contributes to the density of atoms
    out of the condensate mode. This contribution is located inside
    the well and is not plotted here.}
  \label{fig:wavepackets}
\end{figure}

In Fig.~\ref{fig:wavepackets} we show time series of snapshots from
wavepacket simulations. The double peak phenomenon mentioned above is
visible in the $k_0a=11.20$ timeserie. This $k_0$ corresponds to a
crossing of $\dek$ and $\dok$ and thus to both high tranmission and a
marked difference in $\Delta t_e$ and $\Delta t_o$ (cf.
Fig.~\ref{fig:deltas}). 

\subsection{Transmission times}
\label{sec:negative-time-delays}

\begin{figure}[tbp]
  \includegraphics{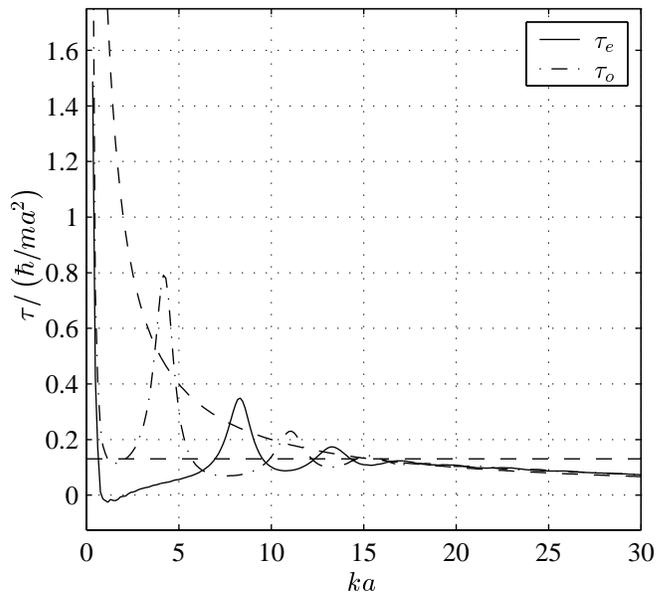}
  \caption{Time spent in the condensate region. Curves are shown for
    both even and odd solutions. Where these curves approximately
    coincide, the retardation can be seen directly in wavepacket
    scattering. When they are very different, the translation to a
    time--dependent wavepacket formulation is less direct. Note the
    negative values of $\tau_e$ in the region around $ka\lesssim 2$:
    Wavepacket simulations show that here the peak of the transmitted
    wavepacket appears \emph{before} the peak of the incident
    wavepacket has reached the condensate. For comparison, the two
    dashed curves show respectively the free motion and the sound wave
    tranmission times, $2a/v_g$ and $2a/c$.}
  \label{fig:tau}
\end{figure}

The time--delays of Eqs.~(\ref{eq:delay_even})
and~(\ref{eq:delay_odd}) can also be translated into an effective
transmission time, the time spent traversing the condensate region.
The well has a width $2a$ so the time spent inside the well can
heuristicly be defined as $\tau=\Delta t + 2a/v_g$. In
Fig.~\ref{fig:tau} we plot $\tau_e(k)$ and $\tau_o(k)$. The two curves
agree for high $k$, while for $ka$ less than $\sim 15$, i.e., for
$v_g$ less than the Bogoliubov speed of sound in the homogeneous part
of the condensate, $c=\sqrt{gn_0/m}$, they show alternating
peaks. Such peaks are signatures of resonances where an even or an odd
number of oscillations fit inside the condensate. 

At low $k$--values, we observe that $\tau_e$ becomes negative over a
rather wide range. For wavepackets with momentum components mainly in
this range, a peak in the transmitted wavepacket can appear
\emph{before} the peak of the incident wavepacket has reached the
condensate. This is confirmed by wavepacket simulations.

Negative transmission times is a wave phenomenon, which together with
superluminal propagation has been observed for light propagation
through wave guides and through dispersive atomic media. It is not
suprising that the Bogoliubov-De Gennes equations show similar
effects, and, indeed, Ray Chiao et al \cite{chiao} have suggested a
many-body interference mechanism that could lead to atomic
transmission through condensates with negative transmission times.
From our Bogoliubov analysis it is not clear if the result is due to
this mechanism or if it is more closely related to the time delays,
which may also be observed in normal wave packet tunneling through
barriers~\cite{hartman62}. 

Note also that the analysis of both a real experiment and of
wavepacket simulations is more complicated for massive particles than
for light: As the transmission coefficient depends strongly on $k$,
there is a velocity filter effect, i.e., the transmitted
wavepacket may move at a different speed than the incoming one.

\section{Conclusion}
\label{sec:conclusion}
It is well known that the Bogoliubov treatment leads to 
excited states ranging from phonon like disturbances of
the mean field amplitude at low energies to particle like excitations at
high enough energies. In the present case of a condensate trapped
in a localized potential minimum of finite depth, the eigenspectrum of
low energy excitations corresponds to states which are particle like in some
regions of space and phonon like in others. Like in normal scattering
theory, wave packets are formed as superpositions, and they propagate
as a consenquence of the phase evolution of the energy eigenstates.
In our study, this propagation implies that particles are incident on
the condensate, they propagate as phonons through the condensate,
where they may be reflected back and forth between the condensate
edges, and eventually they reemerge as reflected or transmitted
particles. We have determined the reflection and transmission
probabilities and the phase shifts, which enable us to derive
time dependent results from our stationary formulation.
The main results of this analysis were double peaked distributions,
reflected from the condensate and transmission with negative time delays.
An experimental demonstration of the latter phenomenon would be an
interesting supplement to similar studies for light transmission.

The particles are indistinguishable, and it is hence not meaningful to
say that it is the same particle that emerges after the scattering
process. The process, however, is coherent, and interferometric
experiments should be able to show that coherence is maintained, just
as experiments with quantum correlated atoms should reveal that also
entanglement is faithfully preserved by the intermediate phonon
excitation, in analogy with recent experiments where surface plasmons
propagate quantum correlated photon pairs through sub wavelenght hole
arrays \cite{altewischer02:_plasm_quant_entan}. We imagine that
scattering experiments of the kind analyzed in this paper may be an
ingredient in the study of controlled atomic dynamics, in particular
for atoms and condensates trapped on chip architectures,
\cite{folman00:_contr_cold_atoms_nanof_surfac,cassettari00:_beam_split_guided_atoms,haensel01:_bose}.

\begin{acknowledgments}
  Discussions with Ray Chiao and a copy of \cite{chiao} prior to
  publication are gratefully acknowledged. This work was supported by
  the Danish National Research Foundation through QUANTOP, the Danish
  Quantum Optics Center.
\end{acknowledgments}

\bibliography{scat1}

\end{document}